# Magnetic Damping in Epitaxial Fe Alloyed with Vanadium and Aluminum

David A. Smith[1], Anish Rai[2,3], Youngmin Lim[1], Timothy Hartnett[4], Arjun Sapkota[2,3], Abhishek Srivastava[2,3], Claudia Mewes[2,3], Zijian Jiang[1], Michael Clavel[5], Mantu K. Hudait[5], Dwight D. Viehland[6], Jean J. Heremans[1], Prasanna V. Balachandran[4,7], Tim Mewes[2,3], Satoru Emori[1]

[1]Department of Physics, Virginia Tech, Blacksburg, VA 24061, U.S.A.

[2]Department of Physics and Astronomy, University of Alabama, Tuscaloosa, AL 35487, U.S.A.

[3]Center for Materials for Information Technology (MINT), University of Alabama, Tuscaloosa, AL 35487, U.S.A.

[4]Department of Material Science and Engineering, University of Virginia, Charlottesville, VA 22904, U.S.A.

[5]Department of Electrical and Computer Engineering, Virginia Tech, Blacksburg, VA 24061, U.S.A.

[6]Department of Materials Science and Engineering, Virginia Tech, Blacksburg, VA 24061, U.S.A.

[7]Department of Mechanical and Aerospace Engineering, University of Virginia, Charlottesville, VA 22904, U.S.A.

**To develop low-moment, low-damping metallic ferromagnets for power-efficient spintronic devices, it is crucial to understand how magnetic relaxation is impacted by the addition of nonmagnetic elements. Here, we compare magnetic relaxation in epitaxial Fe films alloyed with light nonmagnetic elements of V and Al. FeV alloys exhibit lower intrinsic damping compared to pure Fe, reduced by nearly a factor of 2, whereas damping in FeAl alloys increases with Al content. Our experimental and computational results indicate that reducing the density of states at the Fermi level, rather than the average atomic number, has a more significant impact in lowering damping in Fe alloyed with light elements. Moreover, FeV is confirmed to exhibit an intrinsic Gilbert damping parameter of $\simeq 0.001$, among the lowest ever reported for ferromagnetic metals.**

## I. INTRODUCTION

The relaxation of magnetization dynamics (e.g., via Gilbert damping) plays important roles in many spintronic applications, including those based on magnetic switching[1,2], domain wall motion[3,4], spin wave propagation[5,6], and superfluid-like spin transport[7,8]. For devices driven by spin-torque precessional dynamics[1,9,10], the critical current density for switching is predicted to scale with the product of the Gilbert damping parameter and the saturation magnetization [2,11]. Thus, it is desirable to engineer magnetic materials that possess both low damping and low moment for energy-efficient operation. While some electrically insulating magnetic oxides have been considered for certain applications[5,12,13], it is essential to engineer low-damping, low-moment *metallic* ferromagnets for robust electrical readout via giant magnetoresistance and tunnel magnetoresistance. Fe is the elemental ferromagnet with the lowest intrinsic Gilbert damping parameter ($\simeq 0.002$)[14,15], albeit with the highest saturation magnetization ($\simeq 2.0$ T).

Recent experiments have reported that Gilbert damping can be further reduced by alloying Fe with Co (also a ferromagnetic element), with $Fe_{75}Co_{25}$ yielding an ultralow intrinsic Gilbert damping parameter of $\simeq 0.001$[16,17]. However, $Fe_{75}Co_{25}$ is close to the top of the Slater-Pauling curve, such that its saturation magnetization is greater than that of Fe by approximately 20 %[18]. There is thus an unmet need to engineer ferromagnetic alloys that simultaneously exhibit lower damping and lower moment than Fe.

A promising approach towards low-damping, low-moment ferromagnetic metals is to introduce *nonmagnetic* elements into Fe. In addition to diluting the magnetic moment, nonmagnetic elements introduced into Fe could influence the spin-orbit coupling strength $\xi$, which underlies spin relaxation via orbital and electronic degrees of freedom[19–21]. Simple atomic physics suggests that $\xi$ is related to the average atomic number <Z> of the alloy so that, conceivably, damping might be lowered by alloying Fe with lighter (lower-Z) elements. Indeed, motivated by the premise of lowering damping through a reduced <Z> and presumably $\xi$, prior experiments have explored Fe thin films alloyed with V[20,22,23], Si[24], and Al[25]. However, the experimentally reported damping parameters for these alloys are often a factor of >2 higher[22,23,25] than the theoretically predicted intrinsic Gilbert damping parameter of $\simeq 0.002$ in Fe[26] and do not exhibit a significant dependence on the alloy composition[20,23,24]. A possible issue is that the reported damping parameters – obtained from the frequency dependence of ferromagnetic resonance (FMR) linewidth with the film magnetized in-plane – may include contributions from non-Gilbert relaxation induced by inhomogeneity and defects (e.g., two-magnon scattering)[27–36], which can be affected by the alloying. Therefore, how Gilbert damping in Fe is impacted by alloying with low-Z elements remains an open question.

Here, we investigate the compositional dependence of magnetic relaxation at room temperature in epitaxial thin films of ferromagnetic FeV and FeAl alloys. Both alloys are crystalline bcc solid solutions and hence constitute excellent model systems. We employ two configurations of FMR measurements to gain complementary insights: (1) FMR with samples magnetized in the film plane (similar to the prior experiments) to derive the "effective" Gilbert damping parameter, $\alpha_{eff}^{IP}$, which is found to include extrinsic magnetic relaxation due to two-magnon scattering, and (2) FMR with samples magnetized perpendicular to the film plane to quantify the intrinsic Gilbert damping parameter, $\alpha_{int}$, which is free of the two-magnon scattering contribution.

Since Al ($Z$ = 13) is a much lighter element than V ($Z$ = 23), we might expect lower magnetic relaxation in FeAl than FeV, if the smaller <$Z$> lowers intrinsic Gilbert damping via reduced ξ. Instead, we find a significant decrease in magnetic relaxation by alloying Fe with V – i.e., yielding an intrinsic Gilbert damping parameter of ≃0.001, on par with the lowest values reported for ferromagnetic metals – whereas damping in FeAl alloys increases with Al content. These experimental results, combined with density functional theory calculations, point to the density of states at the Fermi level $D(E_F)$ as a plausible dominant factor for the lower (higher) Gilbert damping in FeV (FeAl). We thus find that incorporating a low-$Z$ element does not generally lower damping and that, rather, reducing $D(E_F)$ is an effective route for lower damping in Fe alloyed with a nonmagnetic element. Our findings confirm that FeV is an intrinsically ultralow-damping alloy, as theoretically predicted by Mankovsky et al.[26], which also possesses a lower saturation magnetization than Fe and FeCo. The combination of low damping and low moment makes FeV a highly promising material for practical metal-based spintronic applications.

**II. FILM DEPOSITION AND STRUCTURAL PROPERTIES**

Epitaxial $Fe_{100-x}V_x$ and $Fe_{100-x}Al_x$ thin films were grown using dc magnetron sputtering on (001)-oriented MgO substrates. Prior to deposition, the substrates were annealed at 600 ºC for 2 hours[37]. The base pressure prior to deposition was $< 5\times10^{-8}$ Torr, and all films were grown with an Ar pressure of 3 mTorr. Fe and V (Al) 2" targets were dc co-sputtered to deposit $Fe_{100-x}V_x$ ($Fe_{100-x}Al_x$) films at a substrate temperature of 200 ºC. By adjusting the deposition power, we tuned the deposition rate of each material (calibrated by X-ray reflectivity) to achieve the desired atomic percentage *x* of V (Al). All FeV and FeAl films had a thickness of 25 nm, which is well above the thickness regime where interfacial effects dominate[31,38]. The FeV (FeAl) films were capped with 3-nm-thick V (Al) deposited at room temperature to protect against oxidation, yielding a film structure of MgO/$Fe_{100-x}V_x$(25nm)/V(3nm) or MgO/$Fe_{100-x}Al_x$(25nm)/Al(3nm).

We confirmed the epitaxial bcc structure of our thin films using high resolution X-ray diffraction. 2θ-ω scans show only the (002) peak of the film and the (002) and (004) peaks of the substrate, as shown in Figure 1. Rocking curve scans of the film peaks show similar full-width-at-half-maximum values of $\simeq 1.3º$ irrespective of composition. The epitaxial relation between bcc Fe and MgO is well known[16,39]: the bcc film crystal is rotated 45º with respect to the substrate crystal, such that the [100] axis of the film lies parallel to the [110] axis of the substrate. The absence of the (001) film peak indicates that our epitaxial FeV and FeAl films are solid solutions rather than B2-ordered compounds[40].

## III. MAGNETIC RELAXATION

### 3.1. In-Plane Ferromagnetic Resonance

Many spintronic devices driven by precessional magnetization dynamics are based on in-plane magnetized thin films. The equilibrium magnetization also lies in-plane for soft ferromagnetic thin films dominated by shape anisotropy (i.e., negligible perpendicular magnetic anisotropy), as is the case for our epitaxial FeV and FeAl films. We therefore first discuss FMR results with films magnetized in-plane. The in-plane FMR results further provide a basis for comparison with previous studies[20,22,23,25].

Samples were placed with the film side facing a coplanar waveguide (maximum frequency 50 GHz) and magnetized by an external field $H$ (from a conventional electromagnet, maximum field 1.1 T) along the in-plane [100] and [110] axes of the films. Here, unless otherwise stated, we show results for $H \parallel [110]$ of the film. FMR spectra were acquired via field modulation by sweeping $H$ and fixing the microwave excitation frequency.

Exemplary spectra for Fe, $Fe_{80}V_{20}$, and $Fe_{80}Al_{20}$ are shown in Figure 2, where we compare the peak-to-peak linewidths at a microwave excitation frequency of 20 GHz. We see that the linewidth for $Fe_{80}V_{20}$ shows a $\simeq 25$ % reduction compared to Fe. We further note that the linewidth for the $Fe_{80}V_{20}$ sample here is a factor of $\simeq 2$ narrower than that in previously reported FeV[20]; a possible origin of the narrow linewidth is discussed later. In contrast, $Fe_{80}Al_{20}$ shows an enhancement in linewidth over Fe, which is contrary to the expectation of lower magnetic relaxation with a lower average atomic number.

The FMR linewidth is generally governed not only by magnetic relaxation, but also by broadening contributions from magnetic inhomogeneities[28,41,42]. To disentangle the magnetic relaxation and inhomogeneous broadening contributions to the linewidth, the typical prescription is to fit the frequency $f$ dependence of linewidth $\Delta H_{pp}^{IP}$ with the linear relation[41]

$$\Delta H_{pp}^{IP} = \Delta H_0^{IP} + \frac{h}{g\mu_B\mu_0}\frac{2}{\sqrt{3}}\alpha_{meas}^{IP}f, \quad (1)$$

where $h$ is the Planck constant, $\mu_B$ is the Bohr magneton, $\mu_0$ is the permeability of free space, and $g$ is the $g$-factor obtained from the frequency dependence of the resonance field (see Section IV and Supplemental Material). In Eq. (1), the slope is attributed to viscous magnetic damping, captured by the measured damping parameter $\alpha_{meas}^{IP}$, while the zero-frequency linewidth $\Delta H_0^{IP}$ is attributed to inhomogeneous broadening. The fitting with Eq. (1) was carried out for $f \geq 10$ GHz, where $H$ was sufficiently large to saturate the films. As is evident from the results in Figure 3, $Fe_{80}V_{20}$ has lower linewidths across all frequencies and a slightly lower slope, i.e., $\alpha_{meas}^{IP}$. On the other hand, $Fe_{80}Al_{20}$ shows higher linewidths and a higher slope.

The measured viscous damping includes a small contribution from eddy currents, parameterized by $\alpha_{eddy}$ (Supplemental Material), and a contribution due to radiative damping[43], given by $\alpha_{rad}$ (Supplemental Material). Together these contributions make up $\simeq 20$ % of the total $\alpha_{meas}^{IP}$ for pure Fe and decrease in magnitude with increasing V or Al content. We subtract these to obtain the effective in-plane Gilbert damping parameter,

$$\alpha_{eff}^{IP} = \alpha_{meas}^{IP} - \alpha_{eddy} - \alpha_{rad}. \quad (2)$$

As shown in Figure 4a, $\alpha_{eff}^{IP}$ remains either invariant or slightly decreases in $Fe_{100-x}V_x$ up to $x = 25$, whereas we observe a monotonic enhancement of $\alpha_{eff}^{IP}$ with Al content in Figure 4b. These results point to lower (higher) damping in FeV (FeAl) and suggest a factor other than the average atomic number governing magnetic relaxation in these alloys. However, such a conclusion assumes that $\alpha_{eff}^{IP}$ is a reliable measure of intrinsic Gilbert damping. In reality, $\alpha_{eff}^{IP}$ may include a contribution from defect-induced two-magnon scattering[27–31,35,36], a well-known non-Gilbert relaxation mechanism in in-plane magnetized epitaxial films[27,32–34,44]. We show in the next subsection that substantial two-magnon scattering is indeed present in our FeV and FeAl alloy thin films.

Although Eq. (1) is not necessarily the correct framework for quantifying Gilbert damping in in-plane magnetized thin films, we can gain insight into the quality (homogeneity) of the films from $\Delta H_0^{IP}$. For our samples, $\mu_0 \Delta H_0^{IP}$ is below $\approx$ 1 mT (see Figure 4c,d), which implies higher film quality for our FeV samples than previously reported[20]. For example, $Fe_{73}V_{27}$ in Scheck *et al.* exhibits $\mu_0 \Delta H_0^{IP} \simeq 2.8$ mT[20], whereas $Fe_{75}V_{25}$ in our study exhibits $\mu_0 \Delta H_0^{IP} \simeq 0.8$ mT. Although $\alpha_{eff}^{IP}$ is comparable between Scheck *et al.* and our study, the small $\Delta H_0^{IP}$ leads to overall much narrower linewidths in our FeV films (e.g., as shown in Figs. 2 and 3). We speculate that the annealing of the MgO substrate prior to film deposition[37] – a common practice for molecular beam epitaxy – facilitates high-quality epitaxial film growth and hence small $\Delta H_0^{IP}$ even by sputtering.

### 3.2. Out-of-Plane Ferromagnetic Resonance

To quantify intrinsic Gilbert damping, we performed broadband FMR with the film magnetized out-of-plane, which is the configuration that suppresses two-magnon scattering[28–31]. Samples were placed inside a W-band shorted-waveguide spectrometer (frequency range 70-110 GHz) in a superconducting electromagnet that enabled measurements at fields > 4 T. This high field range is well above the shape anisotropy field of ≤2 T for our films and hence sufficient to completely saturate the film out-of-plane.

The absence of two-magnon scattering in broadband out-of-plane FMR allows us to reliably obtain the measured viscous damping parameter $\alpha_{meas}^{OP}$ by fitting the linear frequency dependence of the linewidth $\Delta H_{pp}^{OP}$, as shown in Figure 5, with

$$\Delta H_{pp}^{OP} = \Delta H_0^{OP} + \frac{h}{g\mu_B\mu_0}\frac{2}{\sqrt{3}}\alpha_{meas}^{OP} f. \quad (3)$$

We note that the zero-frequency linewidth for the out-of-plane configuration $\Delta H_0^{OP}$ (Figure 6c,d) is systematically greater than that for the in-plane configuration $\Delta H_0^{IP}$ (Figure 4c,d). Such a trend of $\Delta H_0^{OP} > \Delta H_0^{IP}$, often seen in epitaxial films[15,33,45], may be explained by the stronger contribution of inhomogeneity to the FMR field when the magnetic precessional orbit is circular, as is the case for out-of-plane FMR, compared to the case of the highly elliptical precession in in-plane FMR[41]; however, the detailed mechanisms contributing to the zero-frequency linewidth remain the subject of future work. The larger $\Delta H_0^{OP}$ at high V and Al concentrations may be due to broader distributions of anisotropy fields and saturation magnetization, or the presence of a secondary crystal phase that is below the resolution of our X-ray diffraction results.

The absence of two-magnon scattering in out-of-plane FMR allows us to quantify the intrinsic Gilbert damping parameter,

$$\alpha_{int} = \alpha_{meas}^{OP} - \alpha_{eddy}, \quad (4)$$

by again subtracting the eddy current contribution $\alpha_{eddy}$. Since we utilize a shorted waveguide, the contribution due to radiative damping does not apply.

From the compositional dependence of $\alpha_{int}$ as summarized in Figure 6a[1], a reduction in intrinsic Gilbert damping is evidenced with V alloying. Our observation is in contrast to the previous experiments on FeV alloys[20,22,23] where the reported damping parameters remain >0.002 and depend weakly on the V concentration. In particular, the observed minimum of $\alpha_{int} \simeq 0.001$ at $x \simeq 25\text{-}30$ is approximately half of the lowest Gilbert damping parameter previously reported for FeV[20] and that of pure Fe[15]. The low $\alpha_{int}$ here is also comparable to the lowest damping parameters reported for ferromagnetic metals, such as $Fe_{75}Co_{25}$[16,17] and Heusler compounds[46–48]. Moreover, the reduced intrinsic damping by alloying Fe with V is qualitatively consistent with the computational prediction by Mankovsky *et al.*[26], as shown by the curve in Figure 6a. Our experimental finding therefore confirms that FeV is indeed an intrinsically ultralow-damping ferromagnet that possesses a smaller saturation magnetization than Fe.

---

[1] We were unable to carry out out-of-plane FMR measurements for FeV with x = 20 (Fig. 2(c,d)) as the sample had been severely damaged during transit.

In contrast to the reduction of $\alpha_{int}$ observed in FeV alloys, FeAl shows an increase in intrinsic damping with increasing Al concentration, as seen in Figure 6b. Recalling that Al has an atomic number of $Z = 13$ that is lower than $Z = 23$ for V, this trend clashes with the expectation that lower $<Z>$ reduces the intrinsic Gilbert damping through a reduction of the atomic spin-orbit coupling. Thus, we are required to consider an alternative mechanism to explain the higher (lower) damping in FeAl (FeV), which we discuss further in Section V.

### 3.3. Magnetic Relaxation: Practical Considerations

For both FeV and FeAl alloys, $\alpha_{int}$ derived from out-of-plane FMR (Figure 6a,b) is consistently lower than $\alpha_{eff}^{IP}$ derived from in-plane FMR (Figure 4a,b). This discrepancy between $\alpha_{int}$ and $\alpha_{eff}^{IP}$ implies a two-magnon scattering contribution to magnetic relaxation in the in-plane configuration (Figure 4a,b). For many applications including spin-torque oscillators and magnonic devices, it is crucial to minimize magnetic relaxation in-plane magnetized thin films. While the in-plane magnetic relaxation ($\alpha_{eff}^{IP} \simeq 0.002$) is already quite low for the FeV alloys shown here, the low intrinsic Gilbert damping ($\alpha_{int} \simeq 0.001$) points to the possibility of even lower relaxation and narrower FMR linewidths by minimizing two-magnon scattering and inhomogeneous linewidth broadening. Such ultralow magnetic relaxation in FeV alloy thin films may be achieved by optimizing structural properties through growth conditions[16] or seed layer engineering[49].

While ultralow intrinsic Gilbert damping values have been confirmed in high-quality epitaxial FeV, it would be desirable for device integration to understand how magnetic relaxation in FeV would be impacted by the presence of grain boundaries, i.e. in polycrystalline thin films. Reports on polycrystalline FeCo[49] suggest intrinsic damping values comparable to those seen in epitaxial FeCo[16,17]. While beyond the scope of this study, our future work will explore the possibility of low damping in polycrystalline FeV thin films.

## IV. SPECTROSCOPIC PARAMETERS

The results presented so far reveal that magnetic relaxation is reduced by alloying Fe with V, whereas it is increased by alloying Fe with Al. On the other hand, FeV and FeAl alloys exhibit similar compositional dependence of the spectroscopic parameters: effective magnetization $M_{eff}$ (here, equivalent to saturation magnetization $M_s$), magnetocrystalline anisotropy field $H_k$, and the g-factor $g$ – all of which are quantified by fitting the frequency dependence of resonance field (Supplemental Material). As shown in Figure 7a, there is a systematic reduction in $M_{eff}$ with increasing concentration of V and Al. We also note in Figure 7b a gradual reduction in magnitude of the in-plane cubic anisotropy. Both of these trends are expected as magnetic Fe atoms are replaced with nonmagnetic atoms of V and Al. The reduction of $M_{eff}$ by $\simeq 20\%$ in the ultralow-damping $Fe_{100-x}V_x$ alloys with $x = 25\text{-}30$, compared to pure Fe, is of particular practical interest. The saturation magnetization of these FeV alloys is on par with commonly used soft ferromagnetic alloys (e.g., $Ni_{80}Fe_{20}$[50], CoFeB[51]), but the damping parameter of FeV is several times lower. Further, while FeV and FeCo in the optimal composition window show similarly low intrinsic damping parameters, FeV provides the advantage of lower moment. With the product $\alpha_{int} M_{eff}$ approximately proportional to the critical current density to excite precessional dynamics by spin torque[2,11], FeV is expected to be a superior material platform for low-power spintronic devices.

The *g*-factor $g = 2(1 + \mu_L/\mu_S)$ is related to the orbital moment $\mu_L$ and spin moment $\mu_S$; the deviation from the spin-only value of $g = 2.00$ provides insight into the strength of spin-orbit coupling $\xi$[52]. As seen in Figure 7c, $g$ increases by 1-2% with both V and Al alloying, which suggests that $\xi$ increases slightly with the addition of these low-*Z* elements. This finding verifies that <*Z*> is not necessarily a good predictor of $\xi$ in a solid. Moreover, the higher $g$ for FeV is inconsistent with the scenario for lower damping linked to a reduced spin-orbit coupling. Thus, spin-orbit coupling alone cannot explain the observed behavior of Gilbert damping in Fe alloyed with low-Z elements.

## V. DISCUSSION

In contrast to what has been suggested by prior experimental studies[20,22–25], we have shown that the reduction of average atomic number by alloying with a light element (e.g., Al in this case) does not generally lower the intrinsic Gilbert damping of Fe. A possible source for the qualitatively distinct dependencies of damping on V and Al contents is the density of states at the Fermi level, $D(E_F)$: it has been predicted theoretically that the intrinsic Gilbert damping parameter is reduced with decreasing $D(E_F)$, since $D(E_F)$ governs the availability of states for spin-polarized electrons to scatter into[21,26,53–55]. Such a correlation between lower damping and smaller $D(E_F)$ has been reported by recent experiments on FeCo alloys[17,50], FeRh alloys[40], CoNi alloys[56], and Heusler compounds[46,48,57]. The similarity in the predicted composition dependence of the Gilbert damping parameter for FeCo and FeV[26] suggests that the low damping of FeV may be correlated with reduced $D(E_F)$. However, no prior experiment has corroborated this correlation for FeV or other alloys of Fe and light elements.

We therefore examine whether the lower (higher) damping in FeV (FeAl) compared to Fe can be qualitatively explained by $D(E_F)$. Utilizing the Quantum ESPRESSO[58] package to perform density functional theory calculations (details in Supplemental Material), we calculated the density of states for Fe, $Fe_{81.25}V_{18.75}$, and $Fe_{81.25}Al_{18.75}$. It should be recalled that although FeV and FeAl films measured experimentally here are single-crystalline, they are solid solutions in which V or Al atoms replace Fe atoms at arbitrary bcc lattice sites. Therefore, for each of the binary alloys, we computed 6 distinct atomic configurations in a 2×2×2 supercell, as shown in Figure 8. The spin-split density of states for each unique atomic configuration is indicated by a curve in Figure 9. Here, $D(E_F)$ is the sum of the states for the spin-up and spin-down bands, averaged over results from the 6 distinct atomic configurations.

As summarized in Figure 9 and Table 1, FeV has a smaller $D(E_F)$ than Fe, whereas FeAl has a larger $D(E_F)$. These calculation results confirm a smaller (larger) availability of states for spin-polarized electrons to scatter into in FeV (FeAl), qualitatively consistent with the lower (higher) intrinsic Gilbert damping in FeV (FeAl).

We remark that this correlation between damping and $D(E_F)$ is known to hold particularly well in the limit of low electronic scattering rates $\tau^{-1}$, where *intra*band scattering dominates[21,54]. Gilmore *et al.* have pointed out that at sufficiently high electronic scattering rates, i.e., when $\hbar\tau^{-1}$ is large enough that *inter*band scattering is substantial, the simple correlation between the strength of Gilbert damping and $D(E_F)$ breaks down. It is unclear whether our FeV and FeAl alloy films at room temperature are in the intraband- or interband-dominated regime. Schoen *et al.* have argued that polycrystalline FeCo alloy films – with higher degree of structural disorder and likely higher electronic scattering rates than our epitaxial films – at room temperature are still well within the intraband-dominated regime[17]. On the other hand, a recent temperature-dependent study on epitaxial Fe suggests coexistence of the intraband and interband contributions at room temperature[15]. A consistent explanation for the observed room-temperature intrinsic damping in our alloy films is that the interband contribution depends weakly on alloy composition; it appears reasonable to conclude that $D(E_F)$, primarily through the intraband contribution, governs the difference in intrinsic Gilbert damping among Fe, FeV, and FeAl.

## VI. SUMMARY

We have experimentally investigated magnetic relaxation in epitaxial thin films of Fe alloyed with low-atomic-number nonmagnetic elements V and Al. We observe a reduction in the intrinsic Gilbert damping parameter to $\alpha_{int} \simeq 0.001$ in FeV films, comparable to the lowest-damping ferromagnetic metals reported to date. In contrast, an increase in damping is observed with the addition of Al, demonstrating that a smaller average atomic number does not necessarily lower intrinsic damping in an alloy. Furthermore, our results on FeV and FeAl cannot be explained by the change in spin-orbit coupling through alloying. Instead, we conclude that the density of states at the Fermi level plays a larger role in determining the magnitude of damping in Fe alloyed with lighter elements. Our work also confirms FeV alloys as promising ultralow-damping, low-moment metallic materials for practical power-efficient spin-torque devices.

**Acknowledgements:**

This research was funded in part by 4-VA, a collaborative partnership for advancing the Commonwealth of Virginia, as well as by the ICTAS Junior Faculty Program. D.A.S. acknowledges support of the Virginia Tech Graduate School Doctoral Assistantship. A. Sapkota and C. M. would like to acknowledge support by NSF-CAREER Award No. 1452670, A.R. and T.M. would like to acknowledge support by DARPA TEE Award No. D18AP00011, and A. Srivastava would like to acknowledge support by NASA Award No. CAN80NSSC18M0023.

We thank M.D. Stiles for helpful input regarding intrinsic damping mechanisms in alloys.

The data that support the findings of this study are available from the corresponding author upon reasonable request.

|  | **Number of Spin-Up States ($eV^{-1}$) at $E_F$** | **Number of Spin-Down States ($eV^{-1}$) at $E_F$** |
|---|---|---|
| Fe | 10.90 | 3.44 |
| $Fe_{81.25}V_{18.75}$ | $6.28 \pm 1.80$ | $4.61 \pm 0.43$ |
| $Fe_{81.25}Al_{18.75}$ | $6.81 \pm 1.58$ | $10.20 \pm 3.03$ |

**Table 1:** Number of spin-up and spin-down states at $E_F$. For $Fe_{81.25}V_{18.75}$ and $Fe_{81.25}Al_{18.75}$, the average and standard deviation of values for the 6 distinct atomic configurations (cf. Figure 8) are shown.

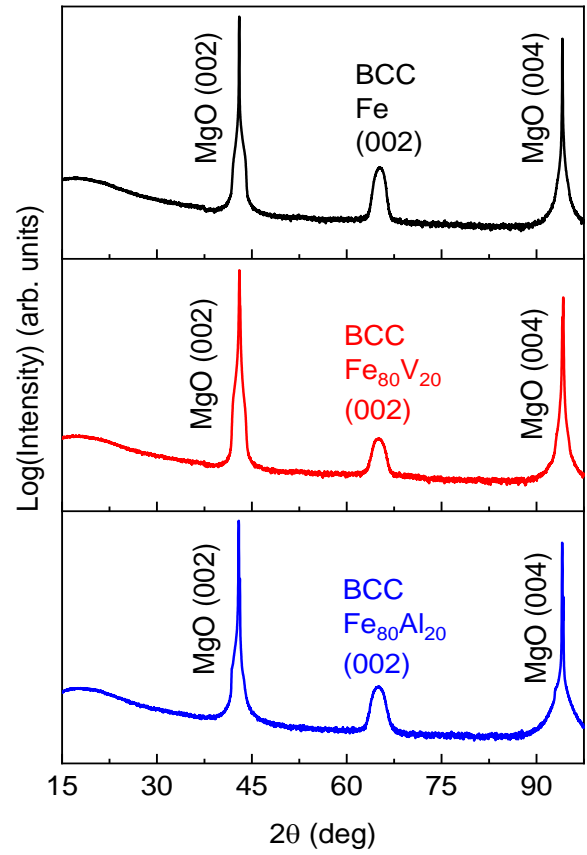

**Figure 1:** (a) 2θ-ω X-ray diffraction scans showing (002) and (004) substrate and (002) film peaks for bcc Fe, $Fe_{80}V_{20}$, and $Fe_{80}Al_{20}$.

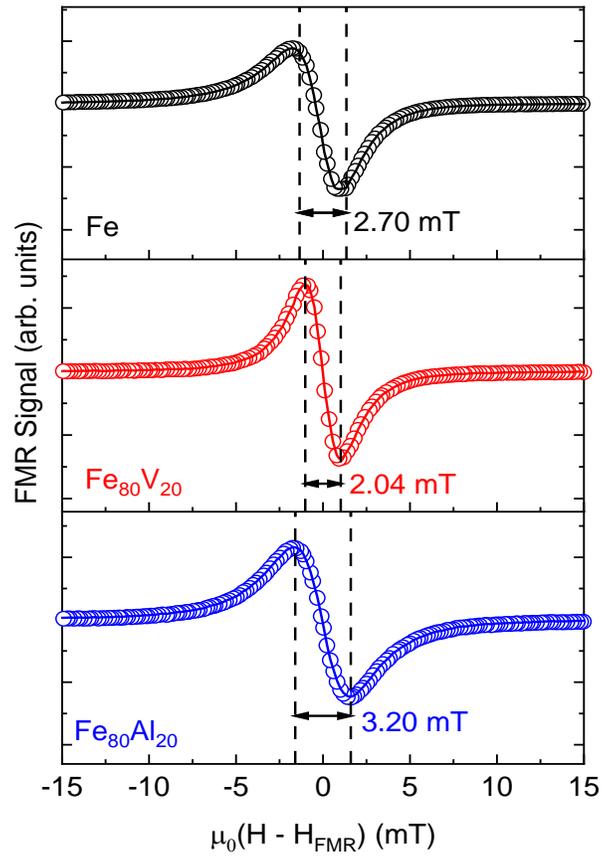

**Figure 2:** FMR spectra at $f = 20$ GHz with the magnetic field $H$ applied in the film plane, fitted using a Lorentzian derivative (solid curve) for Fe, $Fe_{80}V_{20}$ and $Fe_{80}Al_{20}$.

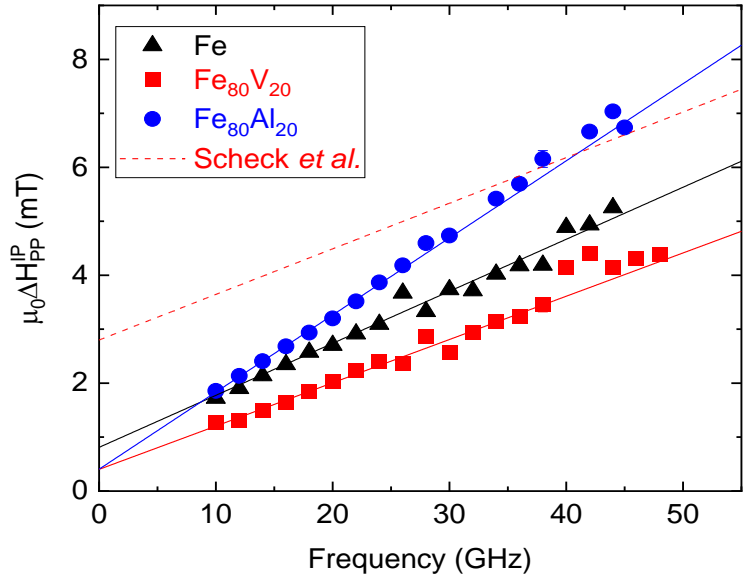

**Figure 3:** FMR linewidths versus microwave frequency for the magnetic field applied within the plane of the film for three distinct alloys. The solid lines are linear fits, described by Eq. (1), from which the effective damping parameter and zero frequency linewidth are determined. The dashed line represents the result for $Fe_{73}V_{27}$ from Scheck *et al.*[20]

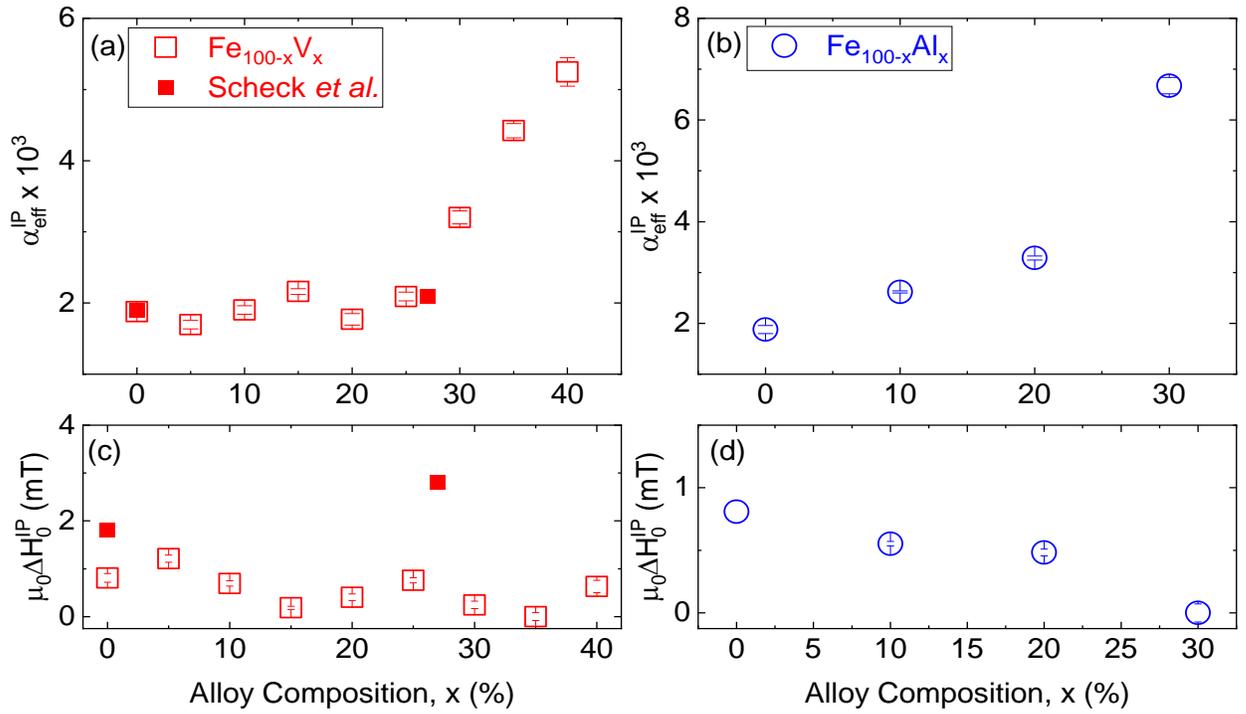

**Figure 4:** The effective damping parameter $\alpha_{eff}^{IP}$ for (a) $Fe_{100-x}V_x$ and (b) $Fe_{100-x}Al_x$ and zero frequency linewidth $\mu_0\Delta H_0^{IP}$ for (c) $Fe_{100-x}V_x$ and (d) $Fe_{100-x}Al_x$, obtained from in-plane FMR. The solid symbols in (a) and (c) represent results reported by Scheck *et al*.[20]

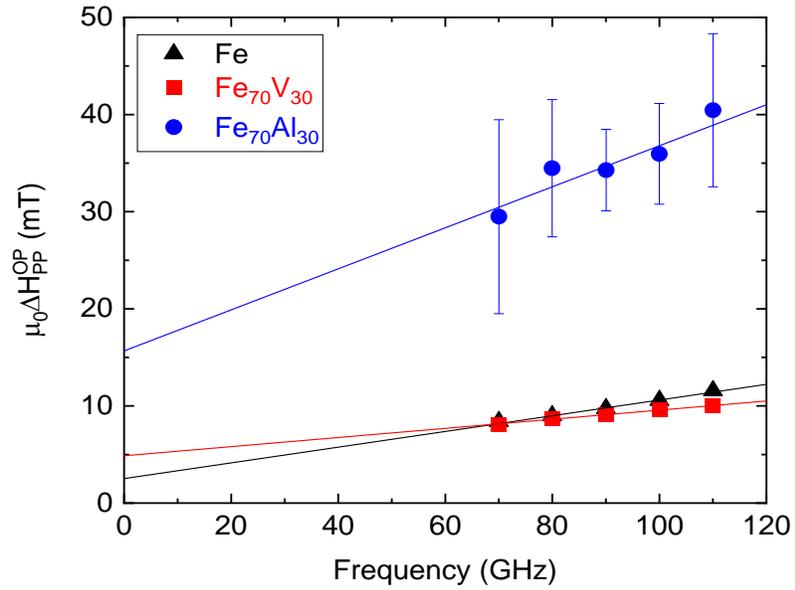

**Figure 5:** FMR linewidths versus applied microwave frequency for the magnetic field applied perpendicular to the plane of the film for three distinct alloys. The line is a linear fit, described by Eq. (3), from which the intrinsic Gilbert damping parameter and zero frequency linewidth are determined.

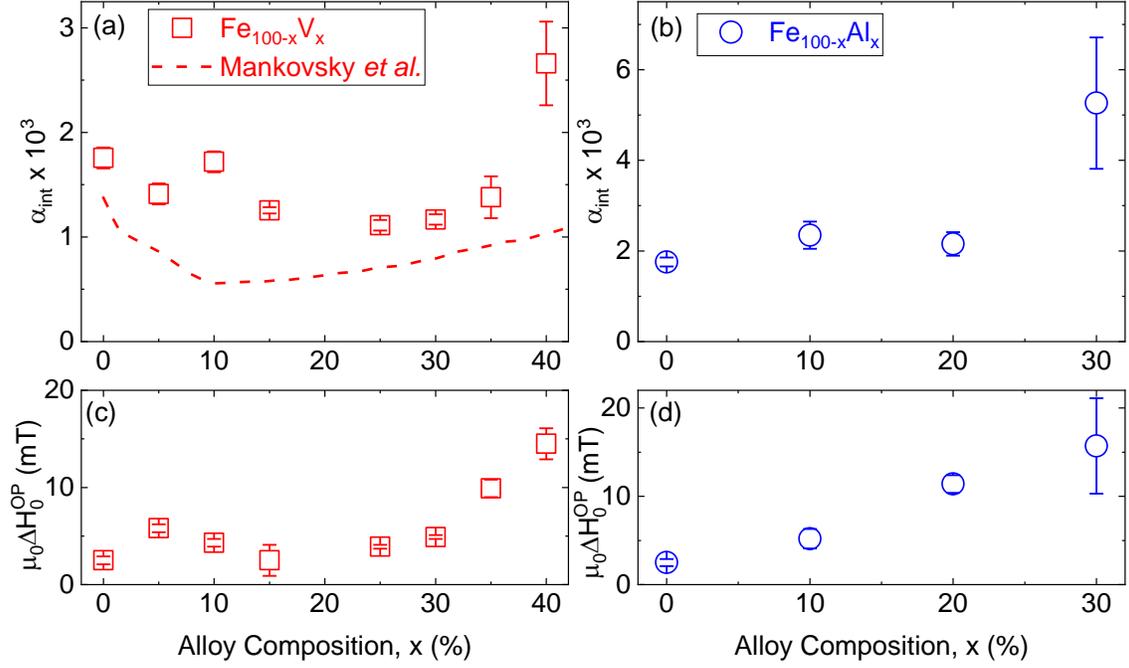

**Figure 6:** The intrinsic Gilbert damping parameter $\alpha_{int}$ for (a) $Fe_{100-x}V_x$ and (b) $Fe_{100-x}Al_x$ and zero frequency linewidth $\mu_0 \Delta H_0^{OP}$ for (c) $Fe_{100-x}V_x$ and (d) $Fe_{100-x}Al_x$, obtained from out-of-plane FMR. In (a), the dashed curve shows the predicted intrinsic damping parameter computed by Mankovsky *et al.*[26]

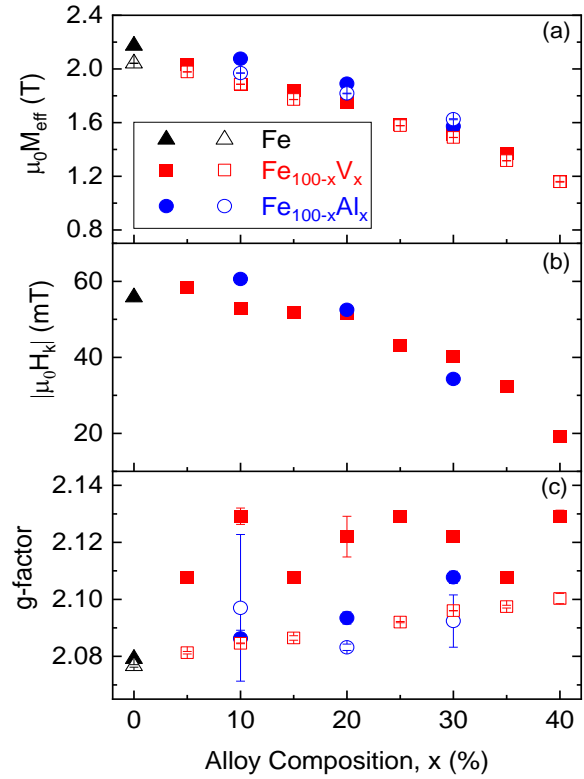

**Figure 7:** (a) Effective magnetization, (b) in-plane cubic anisotropy field, and (c) *g*-factor versus V and Al concentration. The solid (open) markers represent data from in-plane (out-of-plane) measurements.

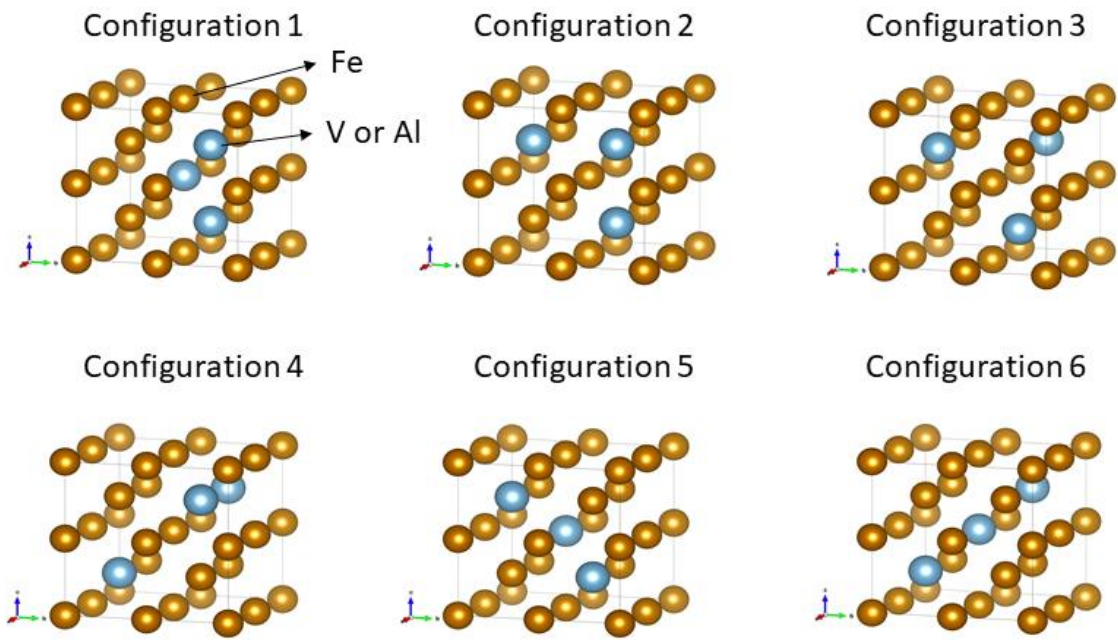

**Figure 8:** The six unique atomic configurations from the supercell program for mimicking the $Fe_{81.25}V_{18.75}$ or $Fe_{81.25}Al_{18.75}$ solid solution.

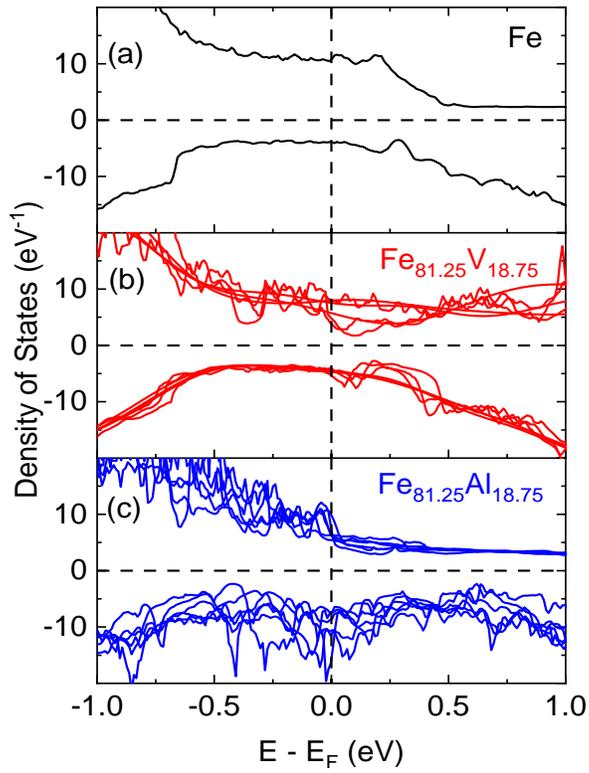

**Figure 9:** Calculated spin-up (positive) and spin-down (negative) densities of states for (a) Fe, (b) Fe$_{81.25}$V$_{18.75}$ and (c) Fe$_{81.25}$Al$_{18.75}$. Results from the 6 distinct atomic configurations are shown in (b,c); the average densities of states at $E_F$ for Fe$_{81.25}$V$_{18.75}$ and Fe$_{81.25}$Al$_{18.75}$ are shown in Table 1.